\documentclass{appolb}
\usepackage{epsfig}% epsfig package included for placing EPS figures in thetext%------------------------------------------------------%%%%%%%%%%%%%%%%%%%%%%%%%%%%%%%%%%%%%%%%%%%%%%%%%%

\begin{document}
% \eqsec 
% uncomment this line to get equations numbered by (sec.num)
\title{Standard Sources of Particle Production in Heavy Ion Collisions}
\author{A. Capella\address{Laboratoire de Physique Th\'eorique (CNRS - UMR 8627),\\
Universit\'e de Paris XI, B\^atiment 210, 91405 Orsay Cedex, France}}
\maketitle
\begin{abstract}
We describe particle production in the framework of an independent string
model~: the dual parton model. We show that an improved version of this model,
containing a diquark breaking component, allows to describe the bulk of particle
production and, in particular, baryon stopping and most of the observed
enhancement of strange baryons. Only for very rare processes, such as $\Omega$
or $J/\psi$ production, the model has to be supplemented with final state
interaction (comovers interaction). Recent data on event-by-event fluctuations
in $p_{T}$ are also described by the model. Predictions for RHIC and LHC are
presented and the effect of nuclear shadowing is discussed. \end{abstract}
%\PACS{PACS numbers come here}  

\vspace{0.5cm}
LPT-ORSAY 99-75

\section{Introduction}
The enhanced production of strange particles (in particular of strange baryons)
and the spectacular suppression of $J/\psi$ in $PbPb$ collisions are considered
by many authors to be signals of Quark Gluon Plasma (QGP) production --or at
least of the production of a system in thermal equilibrium. In this paper, we
describe these phenomena in the framework of the dual parton model (DPM). We
show that the model in its original form fails to describe the large amount of
stopping measured in central $SS$ and $PbPb$ collisions --as do most independent
string models. An improved version of the model containing a diquark breaking
component is introduced. It allows to describe stopping without any extra free
parameter. The same component is also res\-pon\-si\-ble for most of the observed
enhancement of strange baryons. However, the $\Omega$ yield is underestimated
by a factor five. Agreement with experiment is restored by introducing final
state interaction (comovers interaction) with a very small cross-section of
the order of 0.1 mb. Comovers interaction is also needed in order to describe
the anomalous suppression of $J/\psi$ in central $PbPb$ collisions. \par

Comovers interaction is a very non-trivial phenomenon mostly at a partonic
level, which is not entirely understood. It turns out, however, that the
cross-sections required to describe the data, both for strangeness enhancement
and $J/\psi$ suppression are very small and the comovers interaction does not
affect the bulk of particle production. In particular $B\bar{B}$ annihilation
in the final state seems to be negligibly small. \par

Recently, a lot of attention has been devoted to the study of event-by-event
correlations --in particular in $p_{T}$. The small value of this
correlation in central $PbPb$ collisions has been interpreted as a sign of
thermalization. We show that this value is well reproduced in DPM. \par

Predictions of the model for RHIC and LHC are also presented. Although
minijets are important in order to determine $p_{T}$ distributions, in DPM
they do not affect multiplicities. (For the latter, the short $q$-$\bar{q}$
strings in DPM play the same role as minijets). On the contrary, shadowing
corrections are very important and reduce the values of $dN/dy$ at
mid-rapidities by a factor 2 at RHIC and by a factor 3 at LHC. This reduction
is much stronger than in models where minijets are the dominant component. The
reason being that in DPM shadowing corrections are present both for soft and
semi-hard production. 

\section{The model}
The dual parton model (DPM) is a dynamical model for low $p_{T}$ hadronic
and nuclear interactions. It is based on the large-$N$ expansion of
non-perturbative QCD in the Veneziano limit --in which the ratio $N_c/N_f$ is
held fixed \cite{1r}. The dominant configuration consists in the production of
two strings (of type $qq$-$q$ in $pp$ scattering, see Fig. 1). There are also
more complicated configurations, corresponding to higher order terms in the
large-$N$ expansion, involving 4, 6, ... etc strings. These extra strings have
sea quarks and antiquarks at their ends (Fig. 2). These configurations
correspond to multiple inelastic scattering in an $S$-matrix approach. In $pp$
collisions the contribution of each configuration is suppressed by a factor
$N^{-2n+2}$, where $n$ is the number of strings --irrespective of the number of
exchanged gluons and $q$-$\bar{q}$ loops, which do not change the topology of
the graph. It is not known how to compute the numerical values of the various
contributions from the QCD lagrangian. However, Veneziano has shown that there
is a one-to-one correspondence between the various terms in the $1/N$ expansion
and those in a multiple scattering theory (the number of strings being equal to
twice the number of inelastic collisions). In view of that, we determine the
contribution of each configuration to the total cross-section using a multiple
scattering model~: generalized eikonal or perturbative reggeon calculus in
hadron-hadron collisions and Glauber-Gribov model in collisions involving
nuclei. \par

For $A - B$ collisions (for simplicity we consider here $A = B$) in the
approximation of only two strings per nucleon-nucleon collisions, the rapidity
distribution of secondaries is given by \cite{2r}

\begin{equation}
\label{1e}
{dN^{AA} \over dy} (y) = \bar{n}_A \left [ N^{qq_{A_P} - q_{A_T}^v}(y) +
N^{q_{A_P}^v - qq_{A_T}}(y) \right ] + 2(\bar{n} - \bar{n}_A)
N^{q_s - \bar{q}_s}(y) \ . 
\end{equation}

\noindent Here $N(y)$ are the rapidity distributions of the individual strings,
$\bar{n}_A$ is the average number of wounded nucleons of $A$ and $\bar{n}$ is
the average number of collisions. Both $\bar{n}_A$ and $\bar{n}$ can be
computed in the Glauber model. For instance for an average collision (i.e.
integrated over impact parameter), one has

\begin{equation}
\label{2e}
\bar{n} = A^2 \sigma_{NN}/\sigma_{AB} \propto A^{4/3} \ . 
\end{equation}

\noindent Note that the total number of strings is $2\bar{n}$, i.e. two
strings per inelastic nucleon-nucleon collision. \par

The interpretation of (\ref{1e}) is obvious. With $\bar{n}_A$ struck nucleons,
we have at our disposal $\bar{n}_A$ diquarks of projectile and target
$(qq_{A_P}$ and $qq_{A_T}$, respectively) and as many valence quarks. This
accounts for the first term in (\ref{1e}). The remaining strings~: $2(\bar{n} -
\bar{n}_A)$ have to be stretched by sea quarks and antiquarks, because the
available valence constituents are all included in the first term~; this
accounts for the second term of (\ref{1e}). Of course we should combine the
valence and sea constituents of the projectile with those of the target in all
possible ways. However, for linear quantities such as multiplicities, each
ordering gives practically the same result. \par

We can see from (\ref{1e}) and (\ref{2e}) that, if all strings would have the
same plateau height (i.e. the same value of $N(0)$), the plateau height in an
average $AA$ collision would increase like $A^{4/3}$. However, at present
energies, the plateau height of the $q_s$-$\bar{q}_s$, strings is smaller than
that of $qq$-$q$ ones, and the first term of (\ref{1e}) dominates. One obtains
in this way the celebrated wounded nucleon model introduced by our Krakow
hosts \cite{3r}. At higher energies the contribution of the sea strings
becomes increasingly important. Therefore, in order to make predictions for
RHIC and LHC we have to introduce the multistring configurations in each
nucleon-nucleon collision. If their average number is $2\bar{K}$ (this number
can be computed in a generalized eikonal model~: one gets $\bar{K} \simeq 2$
at $\sqrt{s} = 200$ and $\bar{K} \approx 3$ at $\sqrt{s} =$ 7 TeV) the total
number of strings is $2\bar{K}\bar{n}$ and Eq. (\ref{1e}) is changed into

\begin{eqnarray}
\label{3e}
&&{dN^{AA} \over dy} (y) = \bar{n}_A \left [ N^{qq_{A_P} - q_{A_T}^v}(y) +
N^{q_{A_P}^v - qq_{A_T}}(y) + (2 \bar{K} - 2) N^{q_s - \bar{q}_s} \right
] \nonumber \\ &&+ (\bar{n} - \bar{n}_A) 2\bar{K}N^{q_s - \bar{q}_s} \ . 
\end{eqnarray}

The hadronic spectra of the individual strings $N(y)$ is obtained from a
convolution of momentum distribution function and fragmentation functions.
Both are assumed to be universal, i.e. the same in all hadronic and nuclear
interactions. This makes the model very predictive, in particular regarding
the energy and $A$-dependences. Moreover, momentum distribution functions and
fragmentation functions are determined to a large extent from known Regge
intercepts \cite{2r} \cite{4r}. Finally, hadrons produced in different
strings are assumed to be uncorrelated (string independence). This is a
simplicity assumption which does not follow from the large-$N$ expansion. \par

So far, we have only considered configurations with an even number of
strings. What about odd string configurations. A configuration with a single
$3$-$\bar{3}$ string is possible in some cases, i.e. when it is possible to
annihilate a projectile quark with the corresponding antiquark in the target
(Fig. 3). However, in this case flavor quantum numbers are exchanged between
projectile and target and this contribution decreases like $1\sqrt{s}$
(secondary reggeon exchange). \par

A configuration with three $3$-$\bar{3}$ strings is possible in $p\bar{p}$
annihilation (Fig. 4). In this case, baryon number is exchanged between
projectile and target, and the corresponding contribution is also expected to
decrease in $1/\sqrt{s}$. In any case, it is known experimentally that this
contribution is very small at high energy.  

\section{Baryon stopping}
In DPM, the net baryon production takes place from diquark fragmentation. The
produced baryon is fast in average. The data for $pp$ collisions at SPS and for
peripheral $AA$ collisions can be reproduced in this way. However, in the case
of central $SS$ and $PbPb$ collisions, a huge baryon stopping has been observed
by both the NA44 and NA49 collaborations. It cannot be reproduced in the
model. Actually most string models in their original form fail to do so. \par

The origin of this problem resides in the association of the net baryon
production with the diquark. This is not necessarily the case. Indeed, let us
consider again the three string graph for $\bar{p}p$ annihilation of Fig. 4.
Here the valence quarks and antiquarks are found in their corresponding
hemispheres and yet no baryon or antibaryon is present in the final
state. This indicates that baryon number can be independent of valence
quarks. It also shows that it can be transferred over large rapidity distances
and annihilate with the corresponding antibaryon number --very much in the
same way as quark and antiquark annihilate in the one string diagram of Fig. 3.
Assuming that $\sigma_{annihilation} \sim 1/\sqrt{s}$, one obtains for the
rapidity distribution of baryon number (when it is not associated with valence
quarks) $d\sigma/dy \sim \exp (- 1/2 \Delta y)$. \par

The above picture has its theoretical justification in the works of Dosch
\cite{5r} and Rossi and Veneziano \cite{6r}. These authors have constructed a
gauge invariant state vector of the baryon which leads to a picture of the
baryon made out of three quarks bound together by three strings which join in a
point called string junction (SJ). In this picture it is possible to transfer
the SJ over large rapidity distances --leaving the valence quarks behind. In
what follows, this component will be denoted diquark breaking (DB) component -
while the conventional one will be denoted diquark preserving (DP) component.
\par

Recent work based on the above or related ideas can be found in Refs.
\cite{7r}-\cite{13r}. However , in order to understand in this way the huge
stopping observed in central $AA$ collisions, one has to understand why the DB
component des\-cri\-bed above is small in $pp$ and quite large in central $AA$
collisions. In Ref. \cite{10r}, it was argued that the $A$ dependence of the DB
component is stronger than that of the DP one. In this way, a satisfactory
description of the $pp$ and central $SS$ data was obtained --and predictions
for central $PbPb$ turned out to be correct. This was achieved by introducing
one free parameter --which determines the ratio of the DB over DP
contribution in $pp$. However, this approach is not entirely satisfactory
since it requires a sort of fine tuning, namely the size of the DB component
in $pp$ has to be small enough not to spoil the agreement with the data, and
large enough to describe the central $AA$ data thanks to its larger $A$
dependence. \par

In recent works \cite{12r,13r}, we have introduced a different implementation of
the DB mechanism which avoids any fine tuning --as well as the necessity of an
extra parameter. Our first remark is that, although the DB component is
definitely present in the case of a single $pp$ collision, it gives rise to a
three-string configuration (Fig. 5). This is not the dominant, two-string,
configuration in the large-$N$ expansion and, therefore, is expected to be
small. Since its presence is not required to describe the $pp$ data at SPS
energies, we take it equal to zero for simplicity. (Actually any value small
enough to be in agreement with the $pp$ data can be introduced but it will have
practically no effect on the results for central $AB$ collisions). The second
important remark \cite{10r} is that in the case of two inelastic collisions
per nucleon (for instance an incoming proton scattering inelastically with two
nucleons of a nuclear target), the above topological suppression does not
occur. Indeed, in this case the dominant configuration has four strings for
both the DP and the DB components (see Figs. 2 and 6). Therefore a natural
assumption is that, in this case, the two contributions have equal weights
(1/2). The generalization to the case of $n$ elastic collisions per nucleon
is not so obvious. The assumption we have made \cite{12r,13r} (to be checked
by comparing with experiment) is that there is an equal probability $(1/n)$
for the SJ to follow any of the $n$ collisions. In one of them, the SJ will
join a valence diquark and hadronize in the conventional (DP) way. In all
other cases, it will hadronize according to the DB mechanism. \par

With these assumptions, the rapidity distribution of the net baryon $(B -
\bar{B})$ in $AA$ collisions can be written as

\begin{equation}
\label{4e}
{dN^{AA \to B - \bar{B}} \over dy} = {\bar{n}_A \over \bar{n}} \left [ \bar{n}_A
{dN_{DP} \over dy} + (\bar{n} - \bar{n}_A) {dN_{DB} \over dy} \right ] \ .
\end{equation}  

\noindent Here $\bar{n}$ is the average number of binary collisions and
$\bar{n}_A$ the average number of participants of $A$ (for simplicity we
consider only the case $A = B$). The average number of collisions per
wounded (or participant) nucleon in $\bar{n}/\bar{n}_A$. The probability of
the DP component is thus $\bar{n}_A/\bar{n}$ and that of the DB one $1 -
\bar{n}_A/\bar{n} = (\bar{n} - \bar{n}_A)/\bar{n}$. The extra factor
$\bar{n}_A$ in Eq. (\ref{4e}) ensures baryon number conservation (see
below). Note that in the case of a single collision per nucleon ($\bar{n}
= \bar{n}_A$), only the DP component is present. The exact form of
$dN_{DB}/dy$ is given in Refs. \cite{12r,13r}. Its main feature is the
$\exp (-1/2 \Delta y)$ factor, discussed above. In Eq. (\ref{4e}),
$dN_{DP}/dy$ is the conventional diquark fragmentation component. All
relevant formulae are given in Ref. \cite{14r}. Both components are
normalized to 2 (upon integration in $y$). In this way, the net baryon
yield is $2\bar{n}_A$ as required by baryon number conservation. \par

Note that in $pp$ collisions at high energy there is also more than one
collision per nucleon due to unitarity. Therefore, Eq. (\ref{4e}) gives a well
defined prediction for the increase of stopping in hadronic collision at high
energies. An obvious prediction is that the stopping will increase with the event
multiplicity --large event multiplicity corresponding to a large number of
interactions and/or strings. Preliminary data for the $p$-$\bar{p}$ asymmetry in
DIS at HERA are in qualitative agreement with this prediction. \par

The results for the net baryon yield $B - \bar{B}$ in central $SS$ and $PbPb$
collisions are shown in Fig. 7. The $PbPb$ data are well reproduced. The $SS$
ones (which show a larger stopping than in $PbPb$) are not so well described.
Note, however, that the discrepancy between the data and the model predictions
without DB component has been substantially reduced. Note also that there is
no free parameter here.  

\section{Strangeness enhancement}
We have argued in the previous section that in central $AA$ collisions, a large
number of net baryons are produced at mid-rapidities and that they are
dominantly made out of the SJ plus three sea quarks (see Fig. 6). It is then
obvious than a large number of net $\Lambda$, $\Xi$ and $\Omega$ (i.e. an
increase of these yields per participant) will also take place. As a matter of
fact, this is the only possibility to produce net $\Omega$. The experimental
value of the ratio $\bar{\Omega}/\Omega \sim 0.4$ in central $PbPb$ collisions
at mid-rapidities is very much in favor of the above picture. Moreover, there
will also be a substantial increase in the yield of $K^+$ associated to the
production of $\Lambda$'s. \par

In order to obtain the absolute yields of strange baryons one can use trivial
quark counting arguments together with the ratio $S = 2s/(u+d)$ of strange over
non-strange quarks in the sea. Details of the calculations, using values of $S$
in the range $0.2 \div 0.3$, can be found in ref. \cite{13r}. Very similar
results are obtained \cite{12r} by adopting a more phenomenological attitude,
namely, fixing the $\Lambda$, $\Xi$ and $\Omega$ yields at mid-rapidities from
the $pBe$ and $pPb$ data \cite{15r}. Their values in central $AA$ collisions are
then obtained using Eq. (\ref{4e}). \par

Before stating our results it is necessary to recall the mechanism of antibaryon
production in DPM. It consists of two terms. One of them cor\-res\-ponds to the
usual diquark antidiquark pair production in the string brea\-king process. This
term scales with the number of participants. The second one corresponds to the
presence of diquark-antidiquark pairs in the nucleon sea. This term scales with
$\bar{n}-\bar{n}_A$. \par

The corresponding formula (for $A = B$) is

\begin{equation}
\label{5e}
{dN^{AA \to \bar{B}} \over dy} = n_A {dN^{AA \to \bar{B}}_{string} \over dy} +
(\bar{n} - \bar{n}_A) {dN^{AA \to \bar{B}}_{sea} \over dy} \ .  \end{equation}

\noindent For details see \cite{12r,13r} and references therein. \par

Our results are shown in Fig. 8 (dashed lines). The $p$ and $\Lambda$ yields
are well reproduced. The $\Xi$'s are slightly underestimated. However, the
$\Omega$'s are too low by a factor 5. In an attempt to describe the $\Omega$
yield we have introduced the final state interactions~: $\pi N \to K
\Lambda$, $\pi N \to K \Sigma$, $\pi \Lambda \to K \Xi$, $\pi \Sigma \to K
\Xi$ and $\pi \Xi \to K \Omega$, plus the corresponding reactions for the
antiparticles. They are governed by the gain and loss differential equations
\cite{16r}

\begin{equation}
\label{6e}
{dN_i \over d^4x} = \sum_{K, \ell} \sigma_{k\ell} \ \rho_k(x) \ \rho_{\ell}(x) -
\sum_k \sigma_{ik} \ \rho_i(x) \ \rho_k(x) \ . \end{equation}   

\noindent The first term in the r.h.s. of (\ref{6e}) describes the production
of particle $i$ resulting from the interaction of particles $k$ and $\ell$ with
space-time densities $\rho (x)$ and cross-sections $\sigma_{k\ell}$ (averaged
over the momentum distributions of the interacting particles. The second term
describes the loss of particle $i$ due to its interaction with particle $k$.
The initial densities are the ones obtained without final state interaction
and the averaged cross-sections are taken to be the same for all processes.
For details see \cite{12r,13r}. The data are reproduced with a value of the
cross-section as small as 0.14 mb (full lines in Fig. 8). Note that we do not
consider the inverse reactions required by detailed balance. These reactions
give a negligibly small effect since $\rho_K \rho_{\Lambda} \ll \rho_{\pi}
\rho_N$, etc. For the same reason we have neglected strange exchange
reactions such as $\pi \Omega \leftrightarrow \bar{K} \Xi$. Although the
averaged cross-section can be larger than the value used above, this is
overcompensated by the values of the involved densities. Since
$\rho_{\Omega} \ll \rho_{\Xi}$ (by a factor 20 at initial interaction time),
$\pi \Omega \to \bar{K}\Xi$ is disfavored as compared to $\pi \Xi \to
K\Omega$. Likewise, since $\rho_{\bar{K}} \ll \rho_{\pi}$, $\bar{K} \Xi
\to \pi \Omega$ is disfavored as compared to $\pi \Xi \to \bar{K}\Omega$.
\par

Note that final state interaction is by no means a trivial effect. First, it
represents a departure from the idea of independent strings. Second, and
more important, a large contribution to the integrals (\ref{6e}) comes from
interaction times of a few fermi, close to initial time where the system is
in a dense pre-hadronic state. Actually, Brodsky and Muller \cite{17r}
introduced the concept of comover interaction as a coalescence phenomenon at
the partonic level, in order to describe the final state interaction. It is
therefore clear that a lot of theoretical uncertainty is introduced in this
way. The important result, however, is that the cross-sections required to
describe the data are very small and do not affect the bulk of particle
production. \par

In Fig. 9 we show the ratio $\bar{B}/B$ for $pPb$ and for four centrality bins
in $PbPb$. All these ratios decrease significantly between $pPb$ and central
$SS$ collisions and also between central $SS$ and central $PbPb$. Although this
decrease does not contradict thermal or QGP models (the increase in these ratios
at fixed baryochemical potential can be overcompensated by an increase of the
latter), it is not easy to obtain and most global fits in the framework of
those models do not reproduce it. A more crucial test of the thermal and QGP
models is the ratio of different types of antibaryons. An important result of
our model is a ratio $\bar{\Lambda}/\bar{p}$ significantly smaller than one.
This is in agreement with recent preliminary data on the $\bar{p}$ yield from
the NA49 collaboration \cite{18r} --together with published data on the
$\bar{\Lambda}$ yield by the WA97 one \cite{15r}. This is in contradiction
with the prediction of a $\bar{\Lambda}/\bar{p}$ ratio significantly larger
than one given by J. Rafelski \cite{19r}.\par  

\section{$J/\psi$ 
suppression} \protect{\footnote{This subject has been discussed in
great detail in the lectures by C. Gerschel. The discussion in this
paragraph is very sketchy and assumes that the content of her lectures is
known.}}
The decrease of the ratio $J/\psi$ over DY when the centrality
increases was proposed by Matsui and Satz \cite{20r} as a test of a
deconfining phase transition. Shortly afterwards this decrease was found in
$OU$ collisions by the NA36 collaboration. It has also been found in $SU$ and
$PbPb$ --as well as in $pA$ collisions (see C. Gerschel, these proceedings).
The presence of $J/\psi$ suppression in $pA$ collisions, indicates that
another physical mechanism is at work. Most authors consider that it
consists in the interaction of the pre-resonant $c\bar{c}$ pair with
nucleons of the nucleus. A fit of all existing data gives a value of $6 \div
7$ mb for this absorptive cross-section \cite{21r}. The same mechanism
allows to describe $OU$ and $SU$ collisions but fails to describe the $PbPb$
data --which have a larger (``anomalous'') suppression. The latter has been
interpreted as a sign of a deconfining phase transition \cite{21r}. However,
an alternative explanation has been proposed, based on the idea of comover
interactions ($\pi + J/\psi \to D + \bar{D} + X$) of the same type
introduced in the previous section to describe $\Omega$ enhancement. It has
been shown in \cite{22r} that with a cross-section of 0.6 mb one obtains in
this way a reasonable description of all the data. At a quantitative level,
however, the comover picture tends to slightly overestimate the $J/\psi$
suppression both in central $SU$ and in peripheral $PbPb$ collisions and to
underestimate it slightly in very central $PbPb$ collisions. However, this
disagreement is rather small (about $2\sigma$). Moreover, there are large
uncertainties both in the theory and in the ex\-pe\-ri\-ment. In particular,
recent data on $J/\psi$ production in $pA$ collisions by the E866
collaboration \cite{23r} lead to a smaller value of the absorptive
cross-section. If these data were confirmed, the agreement of the comover
picture would improve. Indeed, by reducing the value of the absorptive
cross-section, there would be ``more room'' for comovers in $SU$ collisions.
\par

The discussion above refers to the true $J/\psi$ over DY data. The NA50
collaboration has also presented the so-called minimum bias (MB) analysis in
$PbPb$ collisions. This refers to the ratio

\begin{equation}
\label{7e}
{J\psi \over DY} = \left ( {J/\psi \over MB} \right )_{exp} \times \left ( {MB
\over DY} \right )_{theory} \ . \end{equation}

\noindent Here $MB$ is the single particle inclusive cross-section for charged
particles for minimum bias events --i.e. without requiring the presence of a
$J/\psi$ or a dimuon pair in the final state. The first factor in the r.h.s. of
Eq. (\ref{7e}) has been measured experimentally, while the second one is
calculated theoretically. The advantage of this procedure is that the statistic
is very high (and the statistical errors very small). However, systematic
errors do not cancel here. \par

The ratio obtained in this way shows no saturation at large $E_{T}$. On the
contrary, it continues to decrease steadily at the highest available values of
$E_{T}$. This feature cannot be reproduced in a comovers approach, at least
in its present version. One has to note, however, that one needs some
theoretical assumptions in order to compute the theoretical ratio $MB/DY$. In
particular one assumes that the tail of the $E_{T}$ distributions of $MB$ and
$DY$ are identical. Since both distributions show a very steep fall off at the
tail, this assumption plays a very crucial role in the determination of that
ratio at very large $E_{T}$. \par

The data \cite{24r} can be reproduced in a deconfining scenario \cite{25r}. One
has to introduce two deconfining phase transitions, a first one for the $\chi$
and $\psi '$ and a second one for the direct $J/\psi$. However, in this
approach, the $<p_{T}>$ of the $J/\psi$ versus $E_{T}$ has a decrease at
large $E_T$ which is not seen in the data. On the contrary, the comovers
scenario gives a saturation at large $E_T$ both for the ratio $J/\psi$ over $DY$
and for the $<p_{T}>$ of the $J/\psi$ versus $E_T$ (see last paper in
\cite{22r}). \par

In conclusion, the present data on $J/\psi$ suppression are very interesting but
their interpretation is not yet established. Hopefully, the forthcoming RHIC
data will allow to clarify the situation (see section 6).

\section{Shadowing corrections and predictions at RHIC and LHC} 
Using Eq. (\ref{3e}) with $K \sim 2$ at RHIC and $K \sim 3$ at LHC we obtain for
central collisions at 7 TeV \cite{26r}

\begin{equation}
\label{8e}
{dN^{SS} \over dy} (y^* \sim 0) \sim 2000 \qquad {dN^{PbPb} \over dy} (y^* \sim
0) \sim 7900 \ . \end{equation}

\noindent These results are obtained without taking into account semi-hard
collisions (minijets). As discussed in the introduction, the latter do not
affect the multiplicities, since the average number of strings is constrained by
unitarity. The fact that some of the $q$-$\bar{q}$ strings can be the result of a
semi-hard gluon-gluon interaction, will affect the intrinsic $p_{T}$ of the
string ends and, thus, the $p_T$ distribution of produced particles. However,
average multiplicities are practically unchanged. \par

As explained in \cite{26r}, the values in (\ref{8e}) are upper limits. A
reduction in these figures is expected from shadowing corrections. In hard
processes, shadowing corrections in the nuclear structure functions are well
known. In our approach, however, these corrections are present irrespective of
whether the process is hard or soft. Moreover, they have the same physical
origin and are governed by the same equations in both cases. \par

The physical origin of shadowing corrections can be traced to the
dif\-fe\-ren\-ce between Glauber model and Gribov field theory. The space-time
picture of the interaction is very different in the two cases. Let us consider
hadron-nucleus collisions. In the Glauber model, we have successive (billiard
ball type of) collisions, while in Gribov theory we have ``parallel'' collisions
of different projectile constituents with target nucleons. A key result of Gribov
\cite{27r} is that the $h-A$ amplitude can, nevertheless, be written as a sum of
multiple-scattering diagrams with elastic intermediate states --which have the
same expressions as in the Glauber model. However, in Gribov theory, there are
extra multiple-scattering diagrams which contain all possible diffractive
excitations of the incoming hadron as intermediate states. At present CERN
energies, these extra diagrams lead to corrections to the Glauber formula of
the order of $10 \div 20$~\% in the total cross-sections. However, their
contribution to $dN/dy$ is much larger and leads to a reduction in the figures
in (\ref{8e}) by about a factor 2 at RHIC and a factor 3 at LHC \cite{28r}.
\par

It is well known that the size of high mass excitations of the initial hadron is
controlled by triple Pomeron couplings. It has been shown in \cite{29r} that the
values of the triple reggeon couplings determined from soft diffraction, allow
also to describe hard diffraction measured at HERA. It is also well known
\cite{30r} that the latter determines the size of shadowing effects in the
nuclear structure functions at low $x$. From the analysis in \cite{29r} it
follows that at a scale $Q^2 \sim 1 \ {\rm GeV}^2$ and $x \sim 10^{-2}$ (the
$x$-value relevant at RHIC) the shadowing in the $Pb$ structure function leads
to its reduction by a factor 0.7. At $x \sim 10^{-3}$ (relevant for LHC) the
corresponding reduction is 0.6. Squaring these values (in order to take into
account shadowing correction in both projectile and target in the case of $PbPb$
collisions), we obtain the reductions by a factor 2 at RHIC and 3 at LHC as
stated above. Similar results are obtained \cite{28r} considering that the
process is soft and computing the modifications to $dN/dy$ at $y^* \sim 0$
resulting from the extra terms in the Gribov theory --using the standard value of
the triple Pomeron coupling. \par

The above considerations show that the average virtualities relevant for the
calculation of the shadowing effects in $dN/dy$ at $y^* \sim 0$ at RHIC and LHC
are of the order of 1 GeV$^2$. In models where the dominant contribution is
semi-hard the relevant average virtualities, obtained from perturbative QCD, are
higher and the shadowing effects significantly smaller. \par

With the above values of $dN/dy$ at $y^* \sim 0$ it is possible to compute the
$J/\psi$ suppression at RHIC and LHC resulting from the two mechanisms
discussed in section 5, namely nuclear shadowing and comover interaction. The
suppression resulting from the former mechanism depends only on the absorptive
cross-section which is expected to be the same at all energies. On the
contrary, the density of comovers depends strongly on energy and, as discussed
above, on the shadowing effects. \par

The ratio of $J/\psi$ over $DY$ in a very central $PbPb$ collision at SPS
energies over the corresponding ratio in $pp$ collisions obtained in Ref.
\cite{22r} is 0.23. (This takes into account both nuclear absorption and
comovers interaction). Without nuclear shadowing, the corresponding ratios at
RHIC and LHC are 0.03 and $10^{-4}$, respectively. When shadowing corrections
are taken into account the corresponding ratios are significantly larger~: 0.11
and 0.02. \par

\section{Event-by-event fluctuations in $p_T$}
Let us define the quantity \cite{31r}

\begin{equation}
\label{9e}
Z = \sum_{i=1}^N z_i
\end{equation}
 
\noindent where $N$ is the total number of particles in a single event and

\begin{equation}
\label{10e}
z_i = p_{T_i} - <p_T> \ .
\end{equation}

\noindent Here $p_{T_i}$ is the $p_T$ of particle $i$ in the event and $<\cdots
>$ denotes the average over all events. The correlation $\phi$ is then defined as

\begin{equation}
\label{11e}
\phi = \sqrt{{<Z^2> \over <N>}} - \sqrt{<z^2>}
\end{equation}
  
\noindent where $<z^2>$ is determined by mixing particles from different events.
In Ref. \cite{31r} a simplistic superposition model was considered in which

\begin{equation}
\label{12e}
{<Z^2>_{AA} \over <N>_{AA}} = {<Z^2>_{NN} \over <N>_{NN}} \ .
\end{equation} 

\noindent In this model $\phi$ is the same in $NN$ and $AA$. Experimentally, it
has been found \cite{32r} that $\phi$ decreases by a factor $3 \div 4$ from $NN$
to central $PbPb$. This decrease has been interpreted in \cite{31r} as a sign of
thermalization (see however Ref. \cite{33r}. \par

It turns out that the decrease of $\phi$ observed experimentally is also
reproduced \cite{34r} in a Monte-Carlo formulation of the QGSM \cite{4r} (a
model which is very close to DPM). The rapidity distribution, multiplicity
distribution and $p_T$ distribution measured experimentally \cite{32r} are also
well reproduced by the model \cite{34r}. The results for $\phi$ are
\cite{34r}~: $\phi_{pp} = 9.0$~MeV, $\phi_{PbPb} = 2.4$~MeV at SPS energies and
$\phi_{pp} = 76$~MeV, $\phi_{PbPb} = 79$~MeV at RHIC. As we see the values of
$\phi$ predicted at RHIC are much larger 
than at SPS and practically equal in $pp$ and
central $PbPb$. 

\section{Conclusions}
We have shown that the main properties of particle production in hadronic
and nuclear collisions can be described in a dynamical string model, the DPM
or QGSM. These models do not incorporate non-standard sources of particle
production --such as the formation of a thermally equilibrated system or a
quark gluon plasma. \par

The huge stopping observed in central $SS$ and $PbPb$ collisions requires a
modification of the model consisting in the introduction of a diquark
brea\-king component. The $PbPb$ data are well reproduced in this way with no
extra parameter and predictions for stopping in other systems are obtained.
Strange baryon enhancement can also be described in the framework of this
improved version of DPM. However, for rare processes such as $\Omega$
production (and to a much lesser extent for $\Xi$ production) one has to
introduce final state interaction (comovers interaction). This mechanism is
also required in order to describe $J/\psi$ suppression. Event-by-event
fluctuations in $p_{T}$ are also well described by the model.

\section*{Figure Captions} 
\begin{description}
\item{\bf Fig. 1 :} Two string diagram in $pp$.
\item{\bf Fig. 2 :} a) Four string diagram in $pp$. b) Four string diagram in
$pA$.
\item{\bf Fig. 3 :} One string diagram in $\bar{p}p$.
\item{\bf Fig. 4 :} Three string diagram in $\bar{p}p$.
\item{\bf Fig. 5 :} Three string diagram for the diquark breaking component in
$pp$.
\item{\bf Fig. 6 :} Four string diagram for the diquark breaking component in
$pA$.
\item{\bf Fig. 7 :} Rapidity distribution of the net baryon number $B -
\bar{B}$ in central $SS$ and $PbPb$ collisions. The full (dotted) line is the
result with (without) the diquark breaking component.
\item{\bf Fig. 8 :} $B + \bar{B}$ yields at mid-rapidities for minimum bias
$pPb$ and central $PbPb$ collisions at SPS energies. Full (dashed) lines are
the results with (without) final state interaction.
\item{\bf Fig. 9 :} Same as Fig. 8 for the ratios $\bar{B}/B$. 
\end{description} 

\newpage

\centerline{\bf Figure 1}
\vspace{1cm}

\begin{center}
\hspace{1.2cm}\epsfig{file=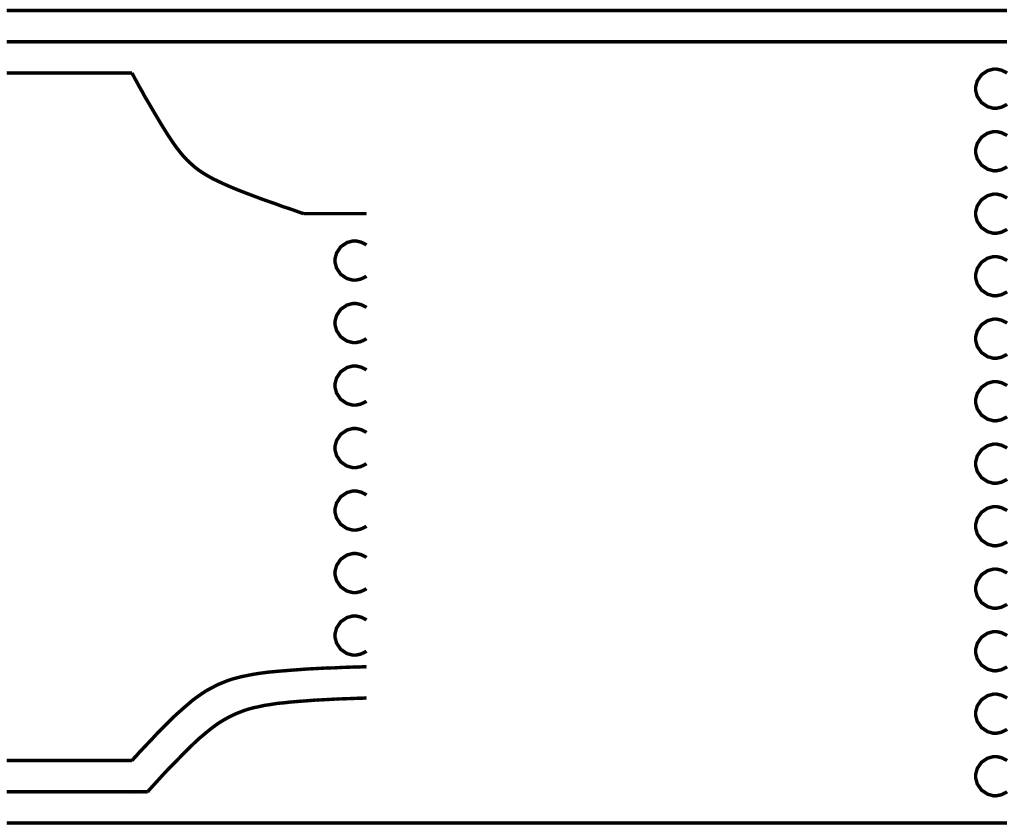,width=8.cm}
\end{center}

\newpage

\centerline{\bf Figure 2}
\vspace{1cm}

\begin{center}
\hspace{1.2cm}\epsfig{file=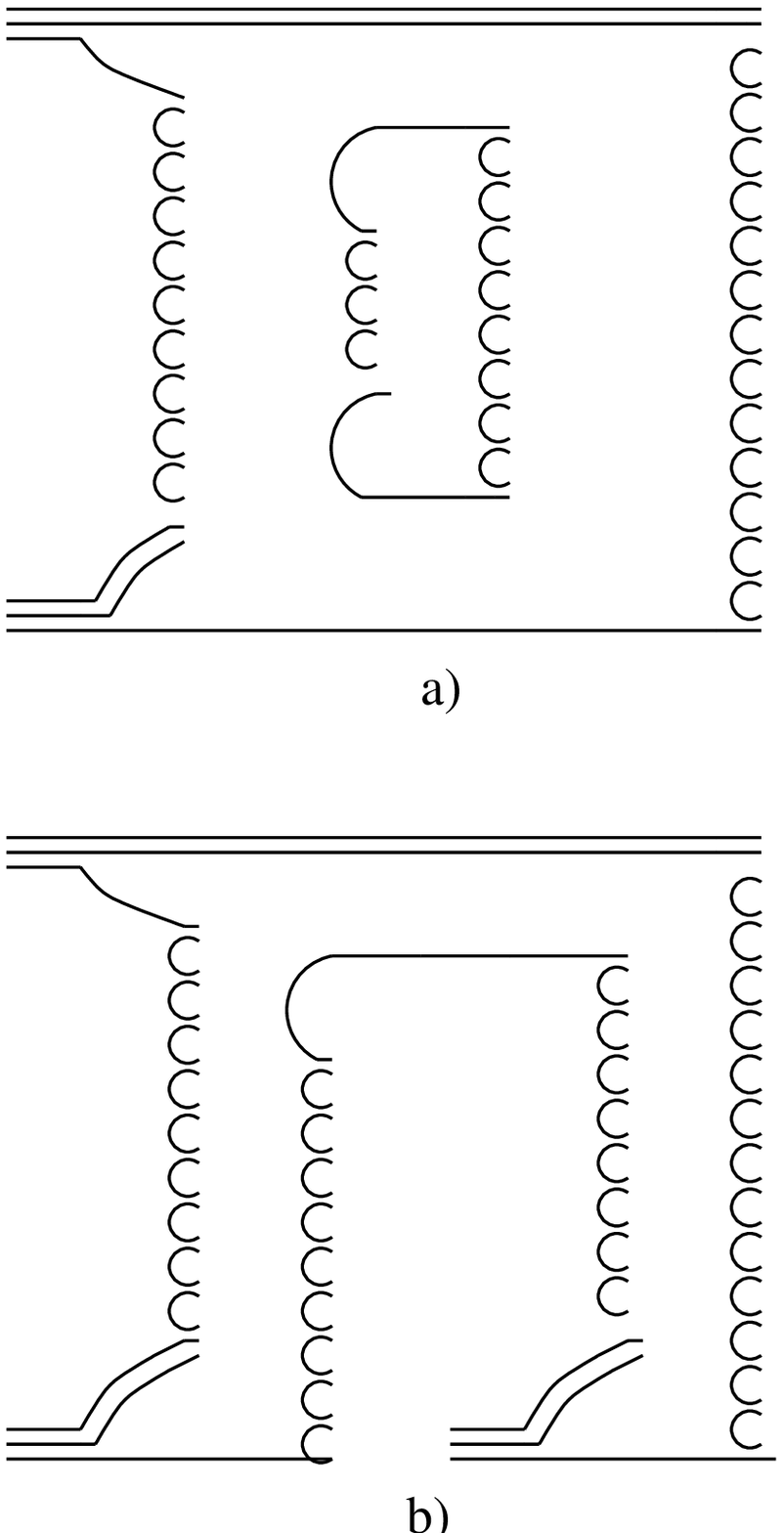,width=8.cm}
\end{center}

\newpage

\centerline{\bf Figure 3}
\vspace{1cm}

\begin{center}
\hspace{1.2cm}\epsfig{file=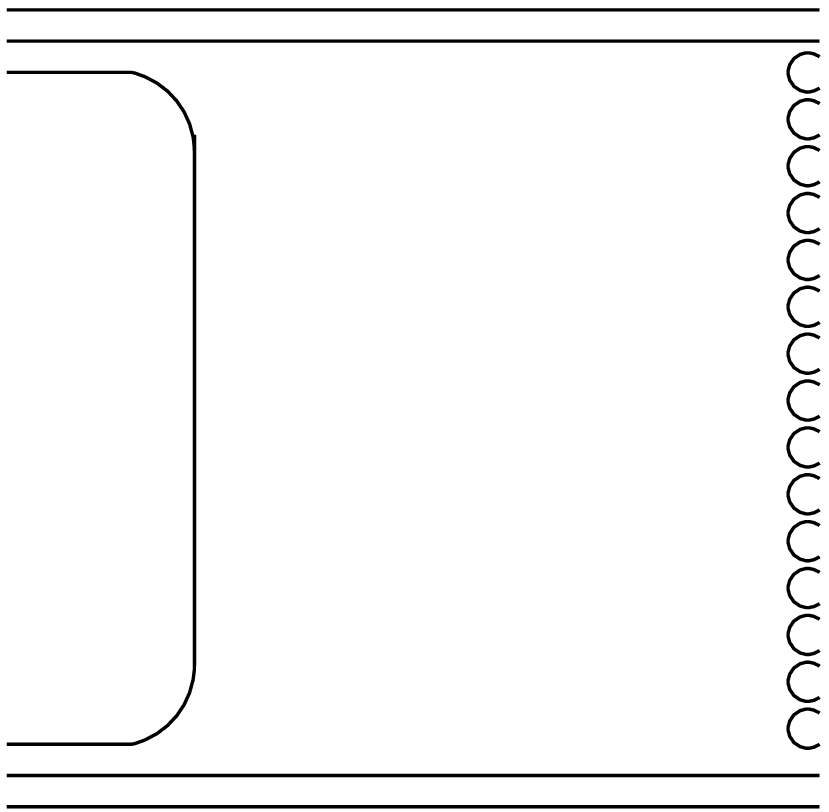,width=8.cm}
\end{center}

\newpage

\centerline{\bf Figure 4}
\vspace{1cm}

\begin{center}
\hspace{1.2cm}\epsfig{file=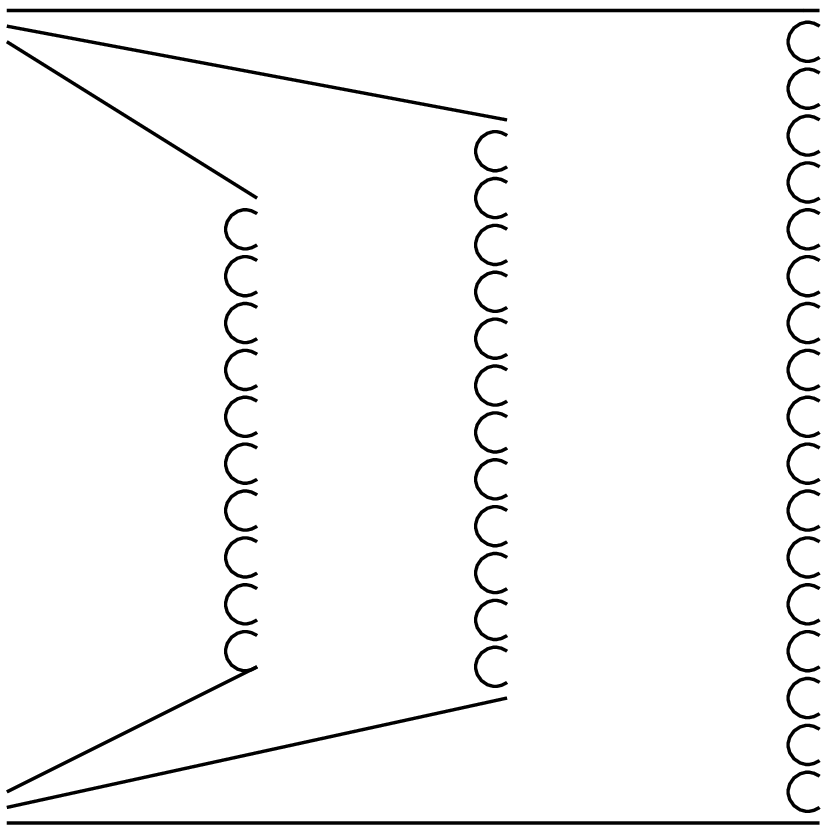,width=8.cm}
\end{center}

\newpage

\centerline{\bf Figure 5}
\vspace{1cm}

\begin{center}
\hspace{1.2cm}\epsfig{file=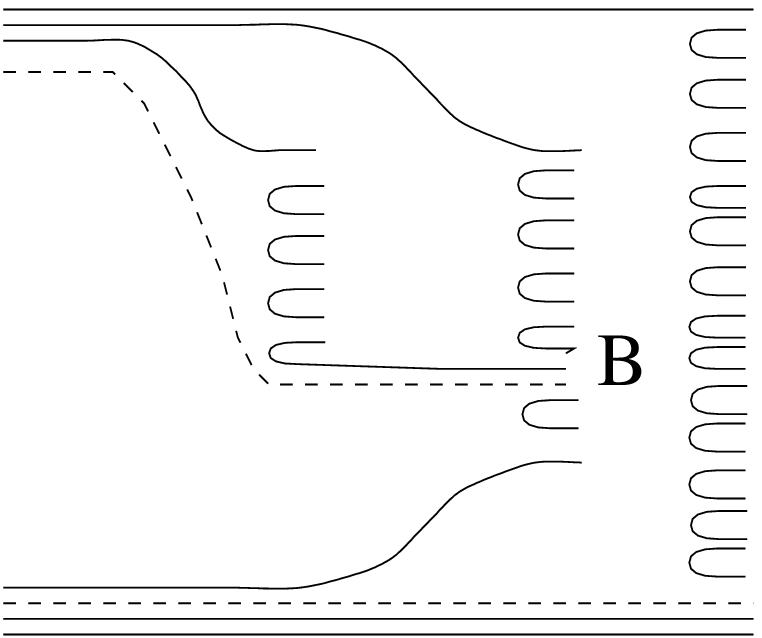,width=8.cm}
\end{center}

\newpage

\centerline{\bf Figure 6}
\vspace{1cm}

\begin{center}
\hspace{1.2cm}\epsfig{file=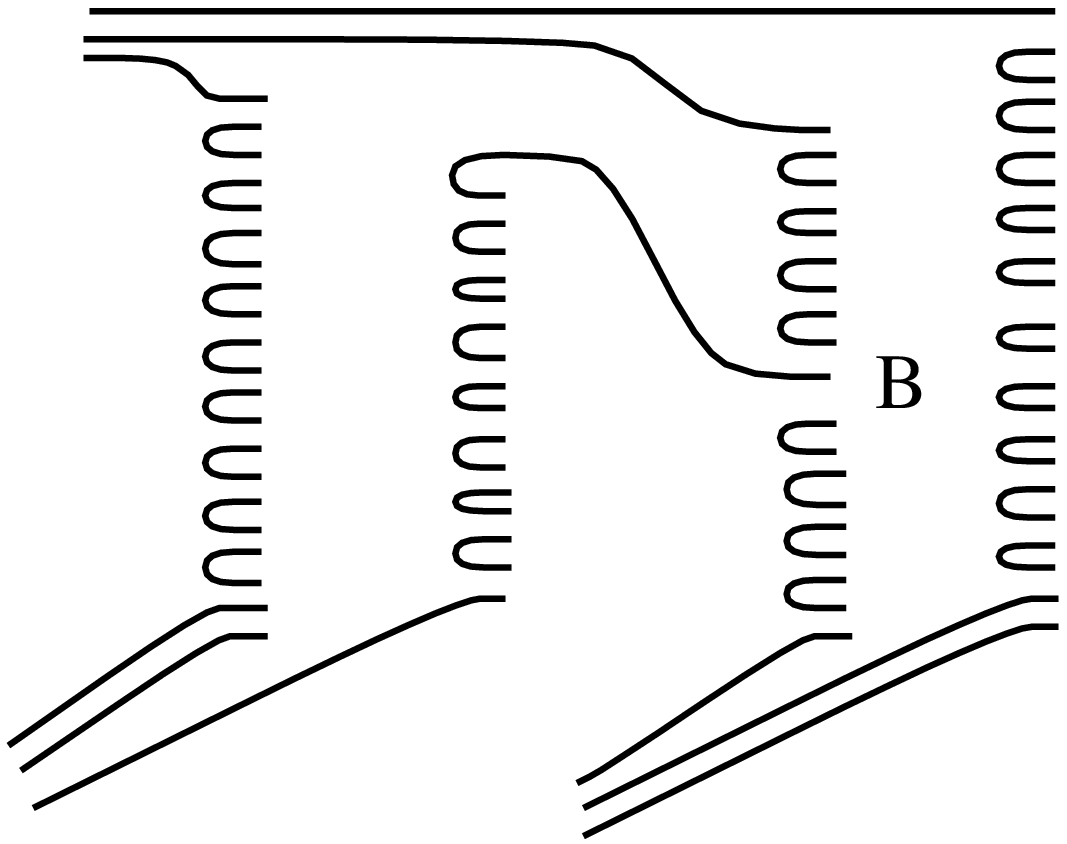,width=9.cm}
\end{center}

\newpage

\centerline{\bf Figure 7}
\vspace{1cm}

\begin{center}
\hspace{1.2cm}\epsfig{file=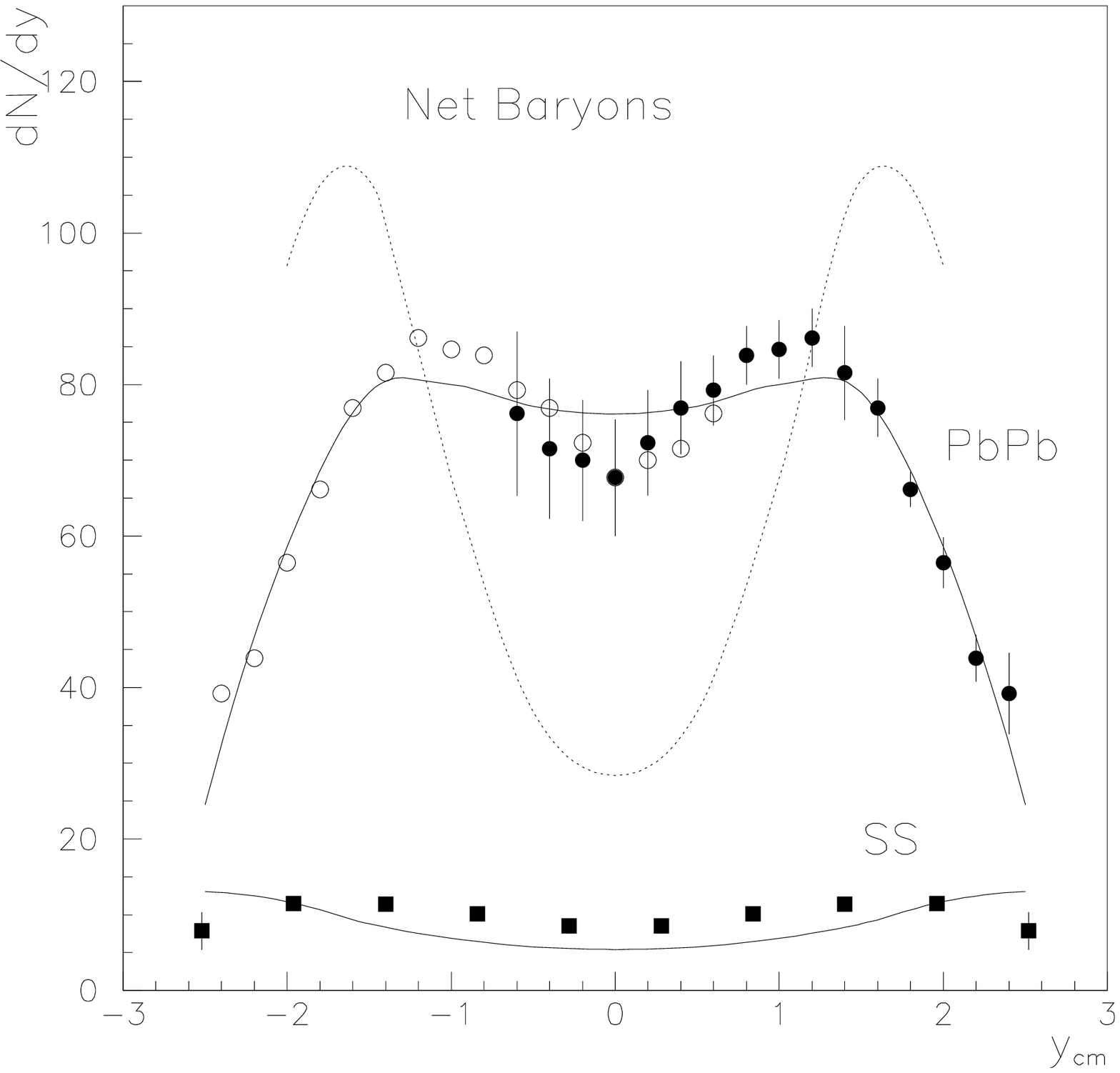,width=14.cm}
\end{center}

\newpage

\centerline{\bf Figure 8}
\vspace{1cm}

\begin{center}
\hspace{1.2cm}\epsfig{file=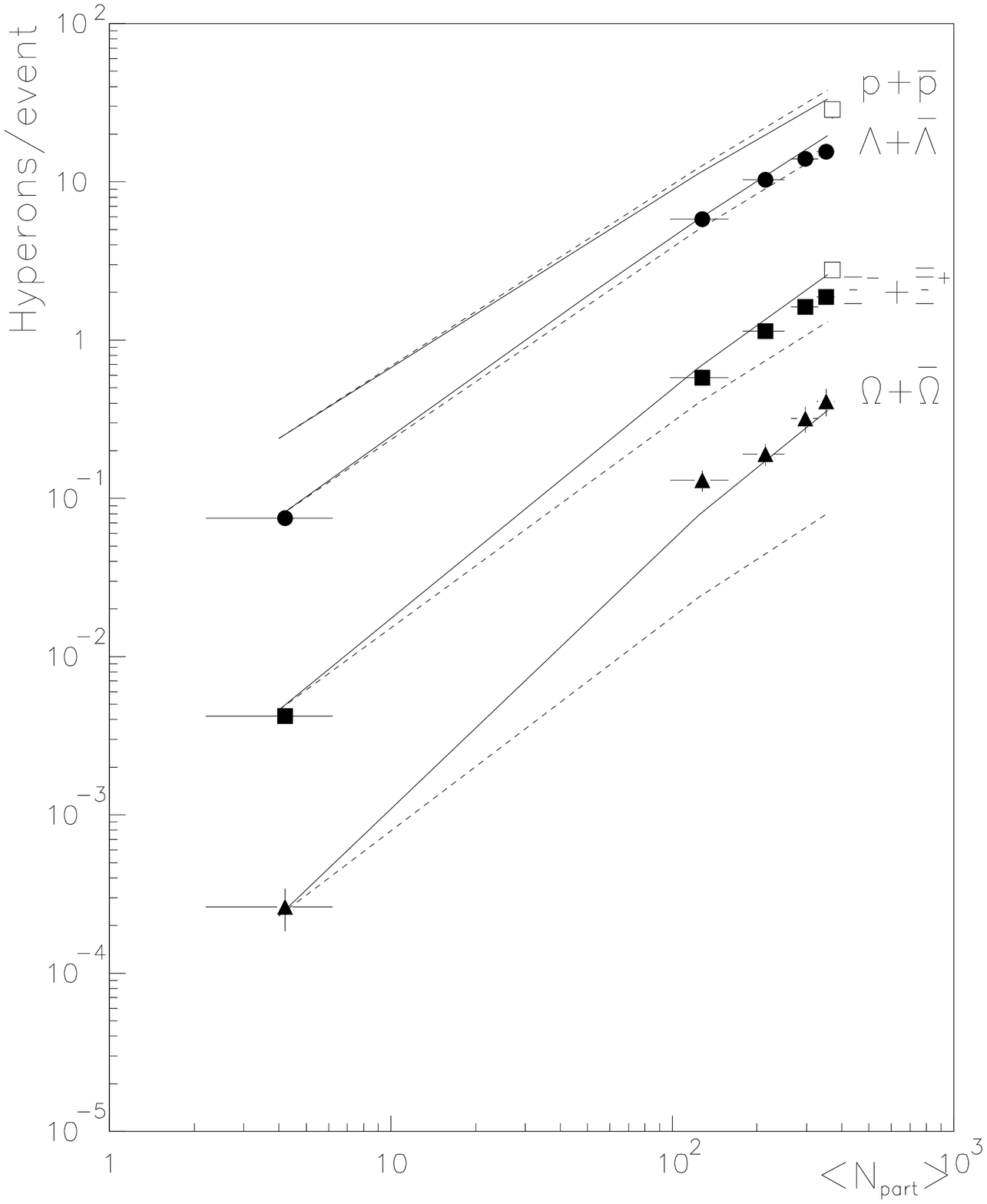,width=12.cm}
\end{center}

\newpage

\centerline{\bf Figure 9}
\vspace{1cm}

\begin{center}
\hspace{1.2cm}\epsfig{file=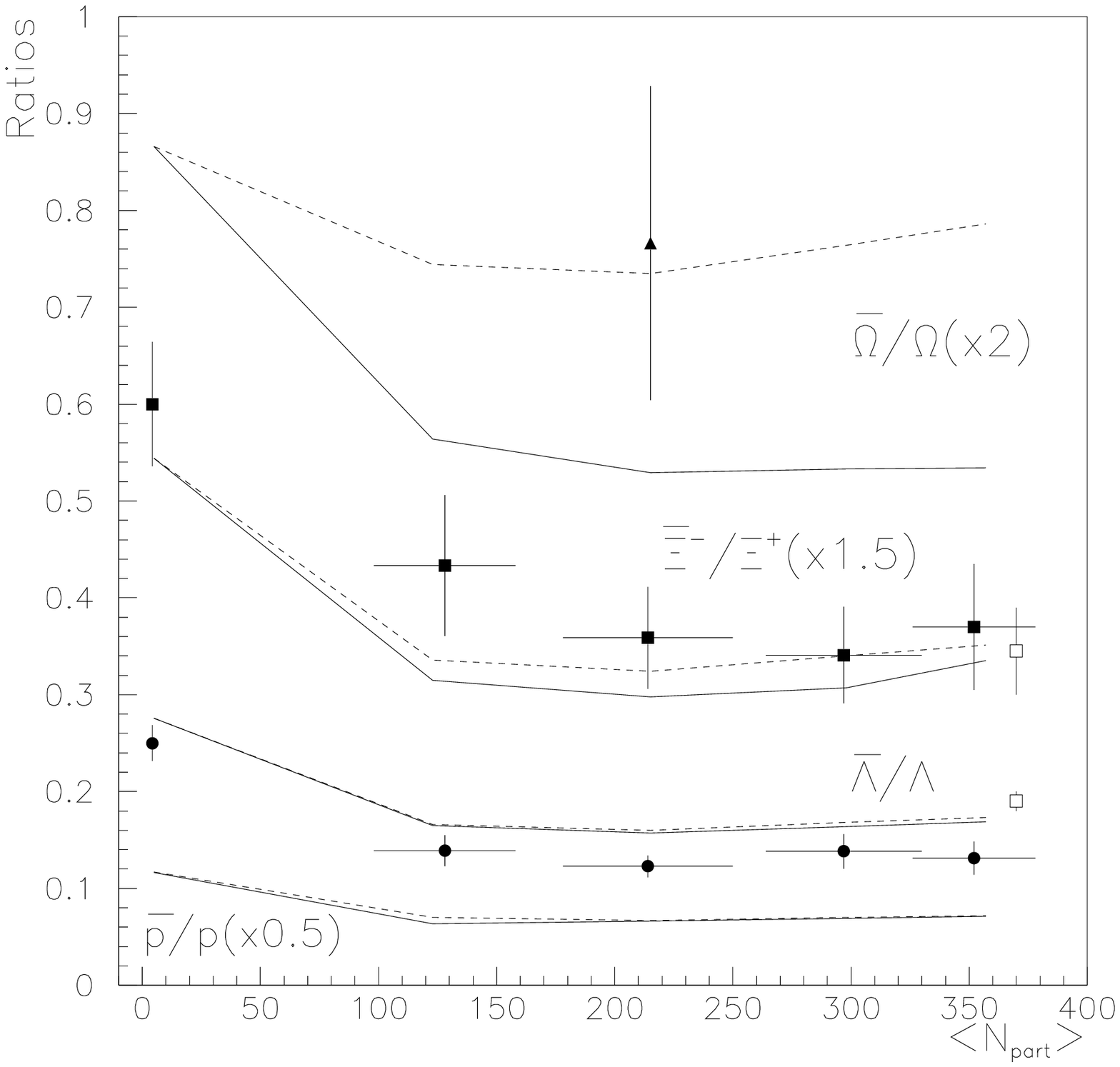,width=13.cm}
\end{center}


\begin{thebibliography}{99}
\bibitem{1r} G. t'Hooft, Nucl. Phys. {\bf B72}, 461 (1974)~;
G. Veneziano, Nucl. Phys. {\bf B74}, 365 (1974). 
\bibitem{2r} DPM : A. Capella, U. Sukhatme, C.
I. Tan and J. Tran Thanh Van, Phys. Lett. {\bf B81}, 69 (1979)~; Phys. Rep. {\bf
236}, 225 (1994). \bibitem{3r} For a review see A. Bialas in Proc. XIIIth Inter.
Symp. on Multiparticle Dynamics, ed. by W. Kittle, W. Metzger and A. Stergiou
(World Scientific 1983).
\bibitem{4r} QGSM~: A. B. Kaidalov, Phys. Lett. {\bf B116}, 459 (1982). A. B.
Kaidalov and K.A A. Ter-Martyrosyan, Phys. Lett. {\bf B117}, 247 (1982).
\bibitem{5r} H. G. Dosch, these proceedings. 
\bibitem{6r} G. C. Rossi and G. Veneziano, Nucl. Phys. {\bf B123}, 507 (1997).
\bibitem{7r} B. Z. Kopeliovich and B. G. Zakharov, Sov. J. Nucl. Phys. {\bf 48},
136 (1988)~; Z. Phys. {\bf C43}, 241 (1989)~; Phys. Lett. {\bf B211}, 221
(1988). 
\bibitem{8r} E. Gotsman and S. Nusinov, Phys. Rev. {\bf D22}, 624 (1980). 
\bibitem{9r} D. Kharzeev, Phys. Lett. {\bf B378}, 238 (1996).
\bibitem{10r} A. Capella and B. Kopeliovich, Phys. Lett. {\bf B381}, 325 (1996).
\bibitem{11r} S. E. Vance and M. Gyulassy, nucl-th/9901009.
\bibitem{12r} A. Capella, E. G. Fereiro and C.A. Salgado, Phys. Lett. {\bf
B459}, 27 (1999).  
\bibitem{13r} A. Capella and C.A Salgado, Phys. Rev. C, in press.
\bibitem{14r} A. Capella, A. B. Kaidalov, A. Kouider-Akil, C. Merino and J. Tran
Thanh Van, Z. Phys. {\bf C70}, 507 (1996).
\bibitem{15r} WA97 collaboration : E. Andersen et al, Phys. Lett. {\bf B433},
209 (1998) and references therein.
 \bibitem{16r} B. Koch, B. Muller and J. Rafelski, Phys. Rep. {\bf 142}, 167
(1986). B. Koch, V. Heinz and J. Pitsut, Phys. Lett. {\bf B243}, 149 (1990).
 \bibitem{17r} S. J. Brodsky and A. H. Mueller, Phys. Lett. {\bf B206}, 685
(1988).
 \bibitem{18r} NA49 collaboration in Proc. Quark Matter 99, Torino, May 1999.
\bibitem{19r} J. Rafelski, these proceedings.
\bibitem{20r} T. Matsui and H. Satz, Phys. Lett. {\bf B178}, 416 (1986). 
\bibitem{21r} D. Kharzeev, C. Louren\c co, M. Nardi and H. Satz, Z. Phys. {\bf
C74}, 307 (1997).
 \bibitem{22r} A. Capella, C. Gerschel and A. Kaidalov, Phys. Lett. {\bf B393},
431 (1997). N. Armesto and A. Capella, Phys. Lett. {\bf B393}, 431 (1997). N.
Armesto, A. Capella and E. G. Ferreiro, Phys. Rev. {\bf C59}, 395 (1999).
\bibitem{23r} E866/Nusea collaboration : M. J. Leitch et al, nucl-exp 9909007.
\bibitem{24r} NA50 collaboration : M. C. Abreu et al, Phys. Rev. {\bf B410}, 327
(1997) and Proc. Quark Matter 99, ibid.
 \bibitem{25r} M. Nardi and H. Satz, Phys. Lett. {\bf B442}, 14 (1998)~; M.
Nardi in Proc. XI Rencontres de Blois, June 1999, Blois, France.
 \bibitem{26r} A.
Capella, C. Merino and J. Tran Thanh Van, Phys. Lett. {\bf B265}, 415 (1991).
 \bibitem{27r} V. N. Gribov, JETP {\bf 56}, 892 (1969)~; JEPT {\bf 57}, 1306
(1969). 
\bibitem{28r} A. Capella, A. Kaidalov and J. Tran Thanh Van, Gribov
Memorial Vo\-lu\-me of Acta Physica Hungarica (Heavy Ion Physics).
 \bibitem{29r} A. Capella, A. Kaidalov, C. Merino, D. Pertermann and J. Tran
Thanh Van, Phys. Rev. {\bf D53}, 2309 (1996). 
\bibitem{30r} A. Capella, A. Kaidalov, C. Merino, D. Pertermann and J. Tran
Thanh Van, Eur. Phys. Journ. {\bf C5}, 111 (1998).
 \bibitem{31r} M. Gazdzicki and St. Mrowczynski, Z. Phys. {\bf C54}, 127 (1992).
\bibitem{32r} NA49 collaboration : H. Appehauser et al., Phys. Lett. {\bf B459},
679 (1999).
 \bibitem{33r} St Mrowczynski, Phys. Lett. {\bf B439}, 6 (1998).
\bibitem{34r} A. Capella, E. G. Ferreiro and A. Kaidalov, hep-ph 9903338, Eur.
Phys. J. C, in press. \end{thebibliography}
\end{document}